\documentclass{vldb}
\usepackage{epsfig} 
\title{S-ToPSS: Semantic Toronto Publish/Subscribe System}

\author{Milenko Petrovic \and Ioana Burcea \and Hans-Arno Jacobsen}

\institution{Department of Electrical and Computer Engineering and Department of Computer Science\\
University of Toronto, Canada\\
\{petrovi, ioana, jacobsen\}@eecg.toronto.edu}

\begin{document}
\maketitle

\section{Introduction}

The increase in the amount of data on the Internet has led to the
development of a new generation of applications based on selective
information dissemination where, data is distributed only to interested
clients. Such applications require a new middleware architecture that can
efficiently match user interests with available information. Middleware
that can satisfy this requirement include event-based architectures such
as publish-subscribe systems (hereafter referred to as pub/sub systems).

The pub/sub paradigm has recently gained a significant interest in
the database community for the support of information dissemination
applications for which other models turned out to be inadequate.

In pub/sub systems, clients are autonomous components that exchange
information by publishing events and by subscribing to the classes
of events they are interested in. In these systems, publishers produce
information, while subscribers consume it. A component usually generates a
message when it wants the external world to know that a certain event has
occurred. All components that have previously expressed their interest in
receiving such events will be notified about it. The central component
of this architecture is the event dispatcher. This component records
all subscriptions in the system. When a certain event is published, the
event dispatcher matches it against all subscriptions in the system. When
the incoming event verifies a subscription, the event dispatcher sends
a notification to the corresponding subscriber.

The earliest pub/sub systems were subject-based. In these systems, each
message (event) belongs to a certain topic. Thus, subscribers express
their interest in a particular subject and they receive all the events
published within that particular subject. The most significant restriction
of these systems is the limited selectivity of subscriptions. The
latest systems are called content-based systems. In these systems,
the subscriptions can contain complex queries on event content.

Pub/sub systems try to solve the problem of selective information
dissemination. Recently, there has been a lot of research on solving
the problem of efficiently matching events against subscriptions
\cite{aguilera99matching,arno}. However, existing matching algorithms
are limited.  For example, if someone is interested in a ``car'', the
system will not return notifications about ``vehicles'' or ``automobiles''
because the matching is based on the syntax and not on the semantics of
the terms.

Matching in a semantic-aware system should correlate the subscriptions and
the publications within a specific knowledge domain. For example, suppose
we have a job-finder application used by interested companies to look
for potential employees. If there is a company recruiter interested in
candidates who graduated from a certain university, with a PhD degree and
with at least 4 years of professional experience, then the recruiter
would subscribe to the following:\\
\centerline{
\textit{\small $S$: (university $=$ Toronto)$\land$(degree $=$ PhD)$\land$}}
\centerline{
\textit{\small (professional experience $\ge$ 4)}}
A prospective candidate would enter the following event using the
job-finder application:\\
\centerline{\textit{\small
$E$: (school, Toronto)(degree, PhD)}}
\centerline{\textit{\small
(work experience, true)(graduation year, 1990)}}
Then the pub/sub system running the job-finder application should
match the event and the subscription above, and send the resume of the
candidate to the company recruiter. Current pub/sub matching algorithms
cannot solve this semantic matching problem.

In this demonstration paper we address the problem of semantic
matching. We investigate how current pub/sub systems can be extended with
semantic capabilities. This is an important issue to be studied because
components in a pub/sub system are a priori decoupled, anonymous, and
do not necessarily ``speak'' the same language.  The main functionality
that a semantic pub/sub system needs to provide is best illustrated
using an example. If a company recruiter is interested in a ``mainframe
developer'', the matching engine should return resumes that not only
contain this exact phrase, but also any resumes that mention ``COBOL
programming'' and years ``1960-1980.''

Our main contribution is the development and validation (through
demonstration) of a semantic pub/sub system prototype S-ToPSS (Semantic
Toronto Publish/Subscribe System).

In the next section we briefly present related work. Section 3 discusses
the S-ToPSS research prototype and its architecture. In Section 4 we
describe the software demonstration.

\section{Related work}

We are not aware of any previous work addressing the semantic matching
problem in pub/sub systems.  Most research on semantic has been done in
the area of heterogeneous database integration. The main problem in this
area is on enabling integration of heterogeneous information systems so
that users can access multiple data sources in an uniform manner. One
way of solving this problem is by using ontologies. Semantic information
systems use an ontology to represent domain-specific knowledge and allow
users to use the ontology terms to construct queries. The query execution
engine accesses the ontology either directly or via an inference engine
in order to optimize the query and generate an execution plan.  Use of an
ontology to generate an execution plan is central in determining the right
source database and method for retrieving the required information. This
allows uniform access to multiple heterogeneous information sources. The
problem of adding semantic capability to pub/sub systems can be seen as an
``inverse'' problem to the heterogeneous database integration problem.
In semantic pub/sub systems, subscriptions are analogous to queries
and events correspond to data, so now the problem is how to match data
to queries.

Some systems \cite{oldsemantic, carnot} use inference engines
to discover semantic relationships between data from ontology
representations. Inference engines usually have specialized languages
for expressing queries different from the language used to retrieve data,
therefore user queries have to be either expressed in or translated into
the language of the inference engine.  The ontology is either global
(i.e.,~domain independent) or domain-specific (i.e.,~only a single
domain) ontology.  Domain-specific ontologies are smaller and more
commonly found than global ontologies because they are easier to specify.
Additionally, there are systems that use mapping functions exclusively and
do not have inference engines \cite{observer,semval}. In these systems,
mapping functions serve the role of an inference engine.

Web service discovery is a process of matching user needs to provided
services; user needs are analogous to events and provided services
to subscriptions in a pub/sub system.  Web service discovery systems
\cite{sem2,sem3} are functionally similar to a pub/sub system. During
a discovery process,  a web service advertises its capabilities in terms
of its inputs and outputs.  An ontology provides an association between
related inputs or outputs of different web services. A user looks for a
particular web service by searching for appropriate inputs and outputs
according to the user's needs. Relevant services are determined by either
exact match of inputs and outputs, or a compatible match according to
ontology relationships.

The main push for using ontologies and semantic information
as means of creating a more sophisticated application
collaboration mechanisms has been from the Semantic Web
community\footnote{www.semanticweb.org}. Recently their focus was on
developing DAML+OIL---a language for expressing, storing and exchange
of ontologies.  Our future work looks at automating translation of
ontologies expressed in DAML+OIL into a more efficient representation
suitable for S-ToPSS.

\section{System architecture}
\subsection{Semantic Event Matching}

In this section we describe how to make the existing matching algorithms
semantic-aware. Our goals are to minimize the changes to the algorithms
so that we can take advantage of their already efficient event
matching techniques and to make the processing of semantic information
fast. We describe three approaches, each adding more extensive semantic
capability to the matching algorithms.  Each of the approaches can be
used independently and for some applications that may be desirable. It
is also possible to use all three approaches together.

The first approach allows a matching algorithm to match events and
subscriptions that use semantically equivalent attributes---synonyms.
The second approach uses additional knowledge about the relationships
(beyond synonyms) between attributes and values to allow additional
matches. More precisely, it  uses a concept hierarchy that provides
two kinds of relations: specialization and generalization.  The third
approach uses mapping functions which allow definitions of arbitrary
relationships between schema and attribute values.

As mentioned earlier, one of the most important features of pub/sub
systems is that the components in a system are decoupled---they are not
aware of each others existence.  Consequently, they do not necessarily use
the same terminology resulting in syntacticly different, but semantically
equivalent schema.  For example, a company recruiter can express her
interest in receiving resumes that match the following constraints:\\
\centerline{\textit{\small
$S$: (university $=$ Toronto)$\land$(professional experience $\ge$ 4)}}
Suppose that there is an applicant's resume with the following:\\
\centerline{\textit{\small
$E$: (school, Toronto)(professional experience, 5)}}
Intuitively, the incoming event should match the subscription. However,
in current pub/sub systems, this will not happen, as ``school'' is not
matched with ``university.'' This exemplifies that syntactic matching
is very limited in the context of current pub/sub systems.

The synonym step involves translating all event and subscription
attributes with different names but with the same meaning, to a ``root''
attribute.  This allows syntactically different event and subscription
attributes to match.  This translation is simple and straightforward,
but the semantic capability it adds to the system may not be sufficient
in some situations, because this approach operates only at attribute
level and does not consider the semantics at the value level within a
predicate (attribute-value pair respectively).  Moreover, this approach
is limited to synonym relations only.

Taxonomies represent a way of organizing ontological knowledge using
specialization and generalization relationships between different
concepts.  Intuitively, all the terms contained in such a taxonomy can
be represented in a hierarchical structure, where more general terms
are higher up in the hierarchy and are linked to more specialized terms
situated lower in the hierarchy.  This structure is called a ``concept
hierarchy.'' Usually, a concept hierarchy contains all terms within a
specific domain, which includes both attributes and values.

Considering the observation that the subscriber should receive only
information that it has precisely requested, we come up with the following
two rules for matching that uses concept hierarchy: (1) the events that
contain more specialized concepts have to match the subscriptions that
contain more generalized terms of the same kind and (2) the events that
contain more generalized terms than those used in the subscriptions do
not match the subscriptions.

For some applications, the semantic functionality obtained by the first
two approaches is sufficient, however, further improvements are possible.
We discussed how to relate semantically identical (synonyms) or similar
(concept hierarchy) information. In both cases, the relationship that was
established between attributes was limited to a single attribute-value
pair only. For example, it may be possible to relate ``university''
and ``school'' as synonyms, but neither the synonym nor the concept
hierarchy can express the relationship between ``graduation year'' and
``professional experience'' as illustrated in the following example.\\
\centerline{\textit{\small
$S$: (university = Toronto)$\land$(professional experience $\ge$ 4)}}
\centerline{\textit{\small
$E$: (school, Toronto)(graduation year, 1993)}}
\centerline{\textit{\small (job1, IBM)(period, 1994-1997)}}
\centerline{\textit{\small (job2, Microsoft)(period, 1999-present)}}
In this resume, the candidate graduated 10 years ago and has had two jobs
since then. Here we have a match between $S$ and $E$ only if we define:\\
\centerline{\textit{\small
professional experience $=$ present date $-$ graduation year}}
This classifies any jobs the potential candidate held in other periods
as not contributing to ``professional experience.''

Mapping functions can specify relationships which otherwise cannot
be specified using a concept hierarchy or a synonym relationship. A
mapping function is a many-to-many function that correlates one
or more attribute-value pairs to one or more semantically related
attribute-value pairs.  It is possible to have many mapping functions
for each attribute. We assume that mapping functions are specified by
domain experts.

\subsection{Semantic Publish/Subscribe System}

\begin{figure}[h]
\centering
\includegraphics[scale=0.5]{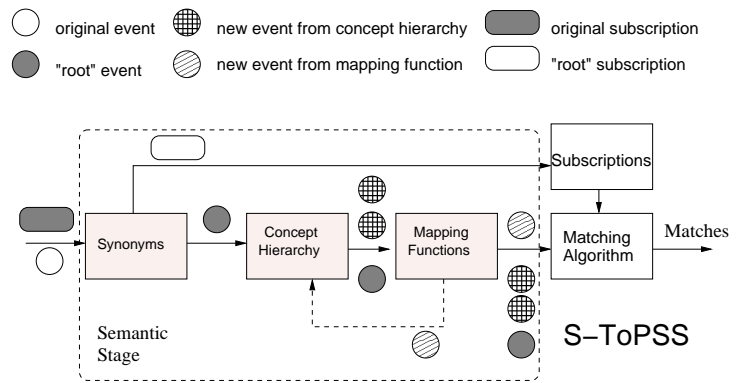} 
\caption{S-ToPSS System Architecture}
\end{figure} 

In this section we describe how the above semantic approaches are combined
to create a fully-fledged semantic pub/sub system.  Figure 1 shows the
S-ToPSS system architecture.  When a new event or a subscription arrives,
the synonym transformation is always done first in order to rewrite
the event/subscription using ``root'' attributes. We can further extend
semantic matching to include more specialized or generalized terms using
a concept hierarchy.  This occurs after the synonym semantic stage.
For each new event, the concept hierarchy stage may create additional
events. The same is true for the mapping function stage. We can see that
mapping function and concept hierarchy stages can be executed multiple
times.  The reason for this is that the concept hierarchy stage can create
new events for which additional mapping functions exist and vice versa.

The main advantage of our approach is performance and flexibility. We
have designed each stage to take advantage of hash structures to quickly
locate relevant information---the key aspect of this approach in terms of
performance---allowing the semantic stage (i.e.,~any combination of the
three stages) to be very fast without affecting already good performance
of the matching algorithms. The flexibility of this approach allows
incremental extension (stage by stage) of matching algorithms, where the
inclusion of any of the three stages improves semantic matching.  It is
also possible to use different semantic stages for different applications.

Furthermore, the use of mapping functions allows a single pub/sub
system to be used for multiple domains simultaneously and, even more
interestingly, it is possible to provide inter-domain mapping by simply
adding additional functions.  This is an very important feature of our
approach, because the current trend is to have many domain-specific
ontologies/concept hierarchies, instead of a single, large and global
ontology. This makes ontology specification easier and more natural.
Thus, being able to use a single pub/sub system for multiple domains
is advantageous.

Some users may be satisfied with fewer results for their semantic
subscriptions, if the matching would be faster. The idea is to allow the
user to inform the system about how much information loss the user is
willing to tolerate. For example, one may only want synonym semantics
to be used or one may restrict the level of a match generality, where
the user is interested only in more general events (e.g.,~a company
recruiter looking to fill an entry-level position would want to receive
resumes from candidates who had some experience with Java, but not from
those who are Java experts).

\section{Software demonstration}

\begin{figure}[h]
\centering
\includegraphics[scale=0.5]{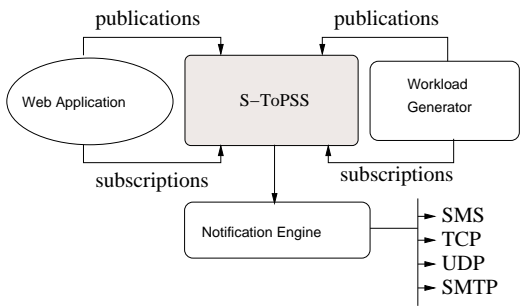} 
\caption{Demonstration Setup}
\end{figure} 

We are going to demonstrate our system using a job-finder application
scenario as an example. In this scenario, we are going to use our system
as an information dissemination service collocated at a job-finder web
server. In this application, companies send subscriptions that specify
qualifications they are looking for from prospective candidates. On the
other hand, candidates send their qualifications as a publication. When a
publication matches a subscription, the candidate's information is sent to
the appropriate company.  The demonstration setup is depicted in Figure 2.

To demonstrate our system, we build a web-based application for client
registration and subscription\slash publication input. We also include a
workload generator that simulates many concurrent clients and companies
sending their subscriptions and publications respectively into the
system. The workload generator creates publications and subscriptions at
random. Moreover, our software demonstration presents a notification
engine that can send notifications to the clients using different
transports.

In order to better understand the advantages of a semantic-aware system,
the application can run in two different modes: semantic or syntactic.
In the semantic mode, the S-ToPSS has all the features as described in
the previous section. In the syntactic mode, only syntax-based matching
is performed.

In conclusion, the real power of this scheme is only apparent by
witnessing how seamlessly unrelated objects end up matching.

\section{Acknowledgements}
We would like to thank Chunhao Yang for his help during the implementation
of the demonstration.

{\small
\bibliography{sempaper}

\begin{thebibliography}{1}

\bibitem{aguilera99matching}
Marcos~Kawazoe Aguilera, Robert~E. Strom, Daniel~C. Sturman, Mark Astley, and
  Tushar~Deepak Chandra.
\newblock Matching events in a content-based subscription system.
\newblock In {\em Symposium on Principles of Distributed Computing}, pages
  53--61, 1999.

\bibitem{oldsemantic}
Y.~Arens and C.~A. Knoblock.
\newblock Planning and reformulating queries for semantically-modeled
  multidatabase systems.
\newblock In {\em Proceedings of the 1st International Conference on
  Information and Knowledge Management}, pages 99--101, 1992.

\bibitem{carnot}
Christine Collet, Michael~N. Hubris, and Wei-Min Sheri.
\newblock Resource integration using a large knowledge base in {CARNOT}.
\newblock {\em IEEE Computer}, pages 55--62, December 1991.

\bibitem{arno}
Francoise Fabret, Arno Jacobsen, Francoise Llirbat, Joao Pereira, Ken Ross, and
  Dennis Shasha.
\newblock Filtering algorithms and implementation for very fast
  publish/subscribe systems.
\newblock In {\em Proceedings of SIGMOD 2001}, 2001.

\bibitem{observer}
E.~Mena, A.~Illarramendi, V.~Kashyap, and A.~P. Sheth.
\newblock {OBSERVER}: An approach for query processing in global information
  systems based on interoperation across pre-existing ontologies.
\newblock {\em International Journal on Distributed and Parallel Databases},
  8(2):223--271, April 2000.

\bibitem{sem2}
Massimo Paolucci, Takahiro Kawamura, Terry~R. Payne, and Katia Sycara.
\newblock Semantic matching of web services capabilities.
\newblock In {\em The 1st International Semantic Web Conference}, 2002.

\bibitem{semval}
Edward Sciore, Michael Siegel, and Arnon Rosenthal.
\newblock Using semantic values to facilitate interoperability among
  heterogeneous information systems.
\newblock {\em ACM Transactions on Database Systems}, 19(2):254--290, 1994.

\bibitem{sem3}
David Trastour, Claudio Bartolini, and Javier Gonzalez-Castillom.
\newblock A semantic web approach to service description for matchmaking of
  services.
\newblock In {\em International Semantic Web Working Symposium}, 2001.

\end{thebibliography}
\bibliographystyle{plain}}
\end{document}